%%%%%%%%%%%%%%%%%% DELTAS0.TEX %%%%%%%%%%%
\input harvmac
\input epsf.tex
\def\caption#1{{\it\centerline{\vbox{\baselineskip=12pt	
\vskip.15in\hsize=4.5in\noindent{#1}\vskip.1in }}}}

\def\dint{\int \kern-.6em€\int\kern-.2em}

\def\({\left(}
\def\){\right)}

\def\too#1{\,\mathop{\longrightarrow}\limits_{#1}\, }
\def\bar#1{\overline{#1}}

 %puts a small half in a displayed eqn
\def\frac#1#2{{\textstyle{#1\over #2}}}

\def\msb{{\mathop{{\bar {MS}}}}}
\def\ltap{\ \raise.3ex\hbox{$<$\kern-.75em\lower1ex\hbox{$\sim$}}\ }
\def\gtap{\ \raise.3ex\hbox{$>$\kern-.75em\lower1ex\hbox{$\sim$}}\ }
\def\gl{\ \raise.5ex\hbox{$>$}\kern-.8em\lower.5ex\hbox{$<$}\ }
\def\roughly#1{\raise.3ex\hbox{$#1$\kern-.75em\lower1ex\hbox{$\sim$}}}

\relax
\def\Dsl{\,\raise.15ex \hbox{/}\mkern-13.5mu D}
\def\[{\left[}
\def\]{\right]}
\def\({\left(}
\def\){\right)}
\def\deldel{\Delta\Delta}
\noblackbox
%\draftmode
%\def\pyidk{PHY-9057135}
\font\ninerm=cmr9
\def\Title#1#2{\nopagenumbers\abstractfont\hsize=\hstitle
\rightline{{\ninerm #1}}
\vskip .4in\centerline{\titlefont #2}
\abstractfont\vskip .5in\pageno=0}
\Title{\vbox{
\hbox{DOE/ER/40427-24-N96}
}}
{\vbox{\centerline{The $\deldel$ Intermediate State in $^1S_0$ $NN$ }
\bigskip
\centerline{Scattering From Effective Field Theory}}}
\vskip-.2in
\centerline{Martin J. Savage\foot{{\tt 
savage@thepub.phys.washington.edu}}}
\centerline{{\sl   Department of Physics, 
University of Washington,
Seattle, WA 98125}}
\bigskip\bigskip\vfill
{We examine the role of the $\deldel$ intermediate state
in NN scattering in the  $^1S_0$ channel. 
The computation is performed at lowest order in an effective field theory 
involving local four-fermion operators and  one pion  
exchange using dimensional regularization 
with $\overline{MS}$.
As first discussed by Weinberg, in the theory with only nucleons
the large scattering length in this channel requires a  small scale for the 
local $N^4$ operators.
When $\Delta$'s are included (but without pions) 
a large scattering length can be obtained 
from  operators with a scale $\sqrt{2 M_N (M_\Delta-M_N)}$, 
but a fine tuning is required.
The coefficients of the contact terms involving the 
$\Delta$ fields are not
uniquely determined but for reasonable values 
one finds that in general $NN$ scattering computed in 
the theory with $\Delta$'s looks like that computed in the 
theory without $\Delta$'s.
The leading effect of the $\Delta$'s is to change the coefficients
of the four-nucleon contact terms between the theories with and without 
$\Delta$'s.
Further, the decoupling of the $\Delta$'s in the limit of large mass
and strong coupling is clearly demonstrated.

When pions are included the typical scale for the contact terms is 
$\sim 100 {\rm MeV}$, 
both with and without $\Delta$'s
and is not set by  $\sqrt{2 M_N (M_\Delta-M_N)}$.
For reasonable values of contact terms that reproduce  
the scattering length and effective range (at lowest order) 
the phase shift is not well reproduced 
over a larger momentum range as is found in the theory without
$\Delta$'s at lowest order.}

\Date{11/96}
\baselineskip 18pt

\newsec{Introduction}

Several years ago Weinberg 
\ref\weinberg{ S. Weinberg, Phys. Lett. B251 (1990) 288;
Nucl. Phys. B363 (1991) 3; Phys. Lett. B295 (1992) 114.}
suggested that low-energy  nucleon-nucleon scattering and 
the interaction of multiple nucleons  ought 
to be described by an effective field theory with a 
systematic expansion in powers of
external momentum and the light quark masses.
For $NN$ scattering at momentum transfers much below the mass 
of the pion the theory is
simply that of four-nucleon operators
and at leading order in the  
expansion one finds that two
such operators are required to describe the S-wave.
The coefficients of these operators are related 
to the scattering lengths in the two spin channels
($S=0$ and $S=1$).
These scattering lengths are much larger than 
what one might expect from the scale of strong interactions due to the 
close proximity  to threshold of a
scattering state in the $S=0$ channel and the deuteron in 
the $S=1$ channel.
Hence, the scales of the coefficients of the operators in the effective 
theory are much smaller than the  QCD scale.
Since Weinberg's work there has been some substantial 
investigations that pursue this approach to nucleon interactions
with the inclusion of pions, operators involving 
the quark masses and operators higher order in the momentum expansion
\ref\bira{C. Ordonez, U. van Kolck, Phys. Lett. B291 (1992) 459;
C. Ordonez, L.Ray,  U. van Kolck, Phys. Rev. Lett. 72(1994) 1982;
Phys. Rev. C53 (1996) 2086.; 
U. van Kolck, Phys. Rev. C49 (1994) 2932.}
\ref\ksw{D. Kaplan, M.J. Savage and M.B. Wise, to appear
in Nuc. Phys. B, (1996), nucl-th/9605002.}.
Further, inelastic processes have been considered 
such as pion-production in NN interactions
\ref\CFMvK{T. Cohen {\it et al}, Phys. Rev. C53 (1996), 2661.}.

It was recently shown how to compute the $^1 S_0$ NN 
scattering amplitude and phase shift including
pion interactions in dimensional regularization
using $\overline{MS}$ to renormalize the
theory \ksw\ , with   
explicit computations   performed at leading and 
subleading order.
Without pions the leading order operators in the 
theory correspond to a zero
range interaction and as a result the effective range 
vanishes.
A non-zero effective range is obtained at next order 
in the expansion and the coefficient of
the operator is fit to reproduce the experimental 
value of $r_{\rm eff.} = 2.73 {\rm fm}$.
When pions are included in the theory they alone 
contribute $1.3 {\rm fm}$ to the effective
range, the rest being made up by a higher dimension 
operator.  
Further, the scale of the coefficients of the local 
four-nucleon operators in the theory with pions 
is substantially
larger than in the theory without pions.
(For comments on this approach see 
\ref\cohen{T. Cohen, nucl-th/9606044 (1996);
D.R. Phillips and T. Cohen, nucl-th/9607048 (1996);
K. A. Scaldeferri {\it et al}, nucl-th/9610049 (1996).}).

The ultimate goal of this line of investigation is 
to describe NN scattering  and inelastic processes involving nucleons
in a theory with systematic power counting.
This means that at a given order in the expansion parameter(s)
an observable can be determined with a given accuracy and that
contributions from higher orders will be parametrically smaller.
One suspects that in order to achieve this 
goal the low-lying baryon resonances
may need to be incorporated as  
we expect the mass difference between the NN and
resonance states will set the scale of the higher dimension operators.
For S-wave scattering, the lowest excited configuration 
that can appear involves the  $N^*(1440)$ and a nucleon.  
Unfortunately, the four baryon operators involving this 
resonance are unknown and are unrelated to quantities we know
by any (approximate) global symmetry.
The next excitation relevent for S-wave scattering is the 
$\deldel$ intermediate state.
Again the four-baryon contact terms have unknown coefficients, 
but the $\Delta$ couplings to pions are reasonably well known 
(satisfying the $SU(4)$ spin-flavor relations at the $10\%$
level, consistent with expectations of large $N_c$ QCD).
As we are 
performing a somewhat exploratory look at the effects of 
higher mass states
we will work with the $\deldel$ intermediate state.
Based on spin and isospin considerations alone there are 
18 four-baryon local operators at lowest
order required to describe the interactions between 
$\Delta$'s and $N$'s.
However, as we are considering NN scattering in the $^1S_0$
channel there are only 3 operators that can
contribute.   
The lagrange density for the contact terms alone is given by
\eqn\locallag{\eqalign{
{\cal L} & = 
-C^{NN} {\cal O}^{NN}
-C^{\Delta N} {\cal O}^{\Delta N}
-C^{\Delta\Delta} {\cal O}^{\Delta\Delta}
}\ \ \ ,}
where
\eqn\localop{\eqalign{
{\cal O}^{NN} & = 
\left[ {1\over 4}\epsilon_{ij} \left( N^{\dagger j}_a 
N^{\dagger i}_b 
+ a\leftrightarrow b\right) \right]
\left[ {1\over 4}\epsilon^{lm} \left( N^a_l N^b_m 
+ a\leftrightarrow b\right) \right]
\cr
{\cal O}^{\Delta N} & = 
\left[ {1\over 4}\epsilon_{ij} 
\left( N^{\dagger j}_c N^{\dagger i}_{c^\prime} 
+ c\leftrightarrow c^\prime\right) \right]
\left[ {3\over 8\sqrt{5}} \epsilon^{i i^\prime}
\epsilon^{j j^\prime}\epsilon^{k k^\prime}
\epsilon_{a a^\prime}\epsilon_{b b^\prime}
\left( \Delta^{a^\prime b^\prime c^\prime}_{i^\prime 
j^\prime k^\prime}
\Delta^{abc}_{ijk} + c\leftrightarrow c^\prime\right)\right]
\cr
{\cal O}^{\Delta \Delta} & = 
\left[ {3\over 8\sqrt{5}} \epsilon_{i i^\prime}
\epsilon_{j j^\prime}\epsilon_{k k^\prime}
\epsilon^{a a^\prime}\epsilon^{b b^\prime}
\left(
\Delta_{  abc}^{\dagger ijk}  
\Delta_{a^\prime b^\prime c^\prime}^{\dagger i^\prime 
j^\prime k^\prime} 
+ c\leftrightarrow c^\prime\right)\right]
\cr
&\ \ 
\left[ {3\over 8\sqrt{5}} \epsilon^{l l^\prime}
\epsilon^{m m^\prime}\epsilon^{n n^\prime}
\epsilon_{d d^\prime}\epsilon_{e e^\prime}
\left( \Delta^{d^\prime e^\prime c^\prime}_{l^\prime 
m^\prime n^\prime}
\Delta^{dec}_{lmn} + c\leftrightarrow c^\prime\right)\right]
}\ \ \ .}
The annihilation operator  $\Delta^{cef}_{lmn}$ for the 
$\Delta$ field 
is symmetric on its
upper flavor indices, each running over $c,e,f  = 1,2$, 
and is symmetric 
on its lower spin
indices which  run over $l,m,n = 1,2$.
Similarly, for the nucleon field $N_i^a$, the upper 
flavor index runs 
over $a=1,2$ and the
lower spin index runs over $i=1,2$.
We have written the operators this way, including their 
constants so that
\eqn\matnorm{\eqalign{
_{a^\prime b^\prime}\langle B^\prime B^\prime ; S=0 | 
{\cal O}^{BB^\prime} | 
BB ; S=0\rangle^{ab}
& =  {1\over 2}\left( 
\delta^a_{a^\prime} \delta^b_{b^\prime} + 
\delta^a_{b^\prime} \delta^b_{a^\prime} 
\right)
}\ \ \  ,}
where $a,b$ and $a^\prime b^\prime$ are isospin indices while 
$B, B^\prime = N $ or $\Delta$.
We see that 
three independent low energy observables are required 
to be measured in order
to fix the  lowest order constants $C^{NN}, C^{\Delta N}$ 
and $C^{\Delta\Delta}$.
Such a requirement is clearly less than desirable, 
providing  a limitation on the
predictive power of the theory when $\Delta$'s are 
included dynamically, and
the situation becomes worse when we consider higher 
dimension operators.
If we are willing to enforce the SU(4) spin-flavor  
symmetry 
that becomes exact in the limit of large $N_c$, 
then the three unknown $C$'s are reduced to two
\ref\kapsavNC{D.B. Kaplan and M.J. Savage, 
Phys. Lett. B365 (1996) 244.}.
How well this symmetry describes the contact interactions of the 
$\Delta$ is still unknown and so we will not use the relations 
between the $C$'s in this work.

At leading order in the momentum expansion there are 
also contributions from pion exchange.
The lagrange density describing such interactions is
\eqn\axiallag{\eqalign{
{\cal L} & = 
-g_A N^\dagger \sigma\cdot A N
- g^{\Delta\Delta} \Delta^\dagger \sigma\cdot A \Delta
- g^{\Delta N} \left( N^\dagger \sigma\cdot A \Delta \ +\ h.c. 
\right)
}\ \ \ ,}
where $A_k$ is the axial-vector field of pions
\eqn\axialdef{\eqalign{
A_k & = {\partial_k\Pi\over f_\pi}
\cr
\Pi & = {1\over\sqrt{2}} \Pi^\alpha\tau^\alpha
}\ \ \ ,}
where $\tau^\alpha$ are the Pauli matrices.
The coupling $g_A=1.25$ is well measured from 
$\beta$-decay, while the other constants are
inferred from the strong decay of the $\Delta$
\ref\jmcoup{E. Jenkins and A.V. Manohar, 
Phys. Lett. B255 (1991) 558;
Phys. Lett. B259 (1991) 353.}
\ref\bssa{M.N. Butler, M.J. Savage 
and R.P. Springer, Nucl. Phys. B399 (1993) 69.}.  
They are found to agree well with the SU(4)
spin-flavour symmetry relations 
\eqn\sufourrel{\eqalign{
{g^{\Delta\Delta} \over g_A} & = {9\over 5}
\ \ \ \  ,\ \ \ \ \ 
{g^{\Delta N}\over g_A}  = {3\sqrt{2}\over 5}
}\ \ \ \  .}

The mass difference between the $\Delta\Delta$ and $NN$ states 
is quite large on the scale of
strong interactions, as we mentioned before, 
and it probably makes little sense to consider a
theory with the $\Delta$'s but without the pions.  
However, the ladder sum of one-pion
exchanges can be made arbitrarily small by allowing 
the axial couplings to become small.   
In this case the leading contribution to the $NN$ 
scattering amplitude is
the sum of bubbles arising from insertions of the local 
four-baryon operators.
We will consider this scenario first as it allows a simple, 
analytical look at the inclusion
of resonances.
We then go on to consider the theory with the true values of 
the axial coupling constants.

\bigskip
\newsec{ The $ g_A = g^{\Delta N} = g^{\Delta\Delta} = 0 $ World}

In the limit of vanishing axial couplings the leading 
contribution to the NN scattering
amplitude is from the sum of all possible chains of 
four-baryon contact terms.
These can be summed explicitly, as was shown by Weinberg 
to be the case for the theory of
nucleons alone.   
It is convenient to write the amplitude for the process as
\eqn\leadnopi{\eqalign{
i {\cal A}_0 & = 
{ -i C_E^*\over 1- C_E^* \ G_E^{NN}({\bf 0},{\bf 0})} }\ \ ,}
where it is straightforward to show that 
\eqn\cstar{\eqalign{
C_E^* = C^{NN} 
+ { (C^{\Delta N})^2 \ G_E^{\Delta\Delta}({\bf 0},{\bf 0})\over 
1- C^{\Delta\Delta}\  G_E^{\Delta\Delta}({\bf 0},{\bf 0}) }
}\ \ \  .}
$G_E^{NN}({\bf 0},{\bf 0})$ and 
$G_E^{\Delta\Delta}({\bf 0},{\bf 0})$ 
are the ${\bf r} = {\bf 0}$ to ${\bf r} = {\bf 0}$
free green functions 
for the $NN$ state and the
$\Delta\Delta$ state of energy $E$ respectively. 
In dimensional regularization they are
\eqn\freegreen{\eqalign{
G_E^{NN}({\bf 0},{\bf 0}) 
& = \int {d^d {\bf q}\over (2\pi)^d} 
{1\over E-{\bf q}^2/M_N + i\epsilon}
\cr
& = -i {M_N |{\bf p}|\over 4\pi} 
+ \ {\cal O}\left( d-3 \right)\ \ \ ,
\cr
G_E^{\Delta\Delta}({\bf 0},{\bf 0}) 
& = \int {d^d {\bf q}\over (2\pi)^d} 
{1\over E-{\bf q}^2/M_N - 2\delta + i\epsilon}
\cr
& = {M_N\over 4\pi}
\sqrt{2 M_N \delta -  |{\bf p}|^2 - i\epsilon }
+\ {\cal O}\left( d-3 \right)
}\ \ \ .}
$\delta = M_\Delta - M_N$ is the $\Delta$-nucleon mass 
difference and for the two nucleon system
we have that $|p|^2 = M_N E$. 
It is straightforward to derive analytic expressions for the
scattering length and effective range in this theory 
(recall that $p \cot \delta  = 
{4\pi\over M_N} Re (1/{\cal A}) = 
-{1\over a} + {1\over 2} r_{\rm eff.} |{\bf p}|^2 + ...$),
\eqn\scattrange{\eqalign{
a & = {M_N\over 4\pi} 
{ C^{NN} + ( (C^{\Delta N})^2  - C^{\Delta\Delta} C^{NN} )\ 
G_{E=0}^{\Delta\Delta}({\bf 0},{\bf 0})
\over 1- C^{\Delta\Delta}\  G_{E=0}^{\Delta\Delta}({\bf 0},{\bf 0})}
\cr
r_{\rm eff} & = 
- {M_N (C^{\Delta N})^2 
\over a^2 4 \pi \ G_{E=0}^{\Delta\Delta}({\bf 0},{\bf 0})}
\left[ {M_N\over 4 \pi} 
{1\over C^{\Delta\Delta}\  G_{E=0}^{\Delta\Delta}({\bf 0},{\bf 0})
-1}\right]^2
 }\ \ \ .}
Notice that while the scattering length can have either sign, 
the effective range is negative
definite.

In the case of $C^{\Delta N}=0$ we return to the 
dynamics discussed by Weinberg where in
order to reproduce the large scattering length in this 
channel of $a_0 = -23.7 {\rm fm}$, the
coefficient $C_E^* = C^{NN} = 4\pi a_0/M_N = -1/(25 {\rm MeV})^2$
is much larger than one would
have guessed from the scale of strong interactions alone.
Unfortunately, we do not know what the coefficients 
$C^{NN}, C^{\Delta N}$ and $C^{\Delta\Delta}$
are, however, it would be unnatural for them to differ 
by orders of magnitude.
One can see from \cstar\ that if 
\eqn\accident{\eqalign{
C^{\Delta\Delta}  \sim  
\left[  G_{E=0}^{\Delta\Delta} ({\bf 0},{\bf 0})  \right]^{-1}
}\ \ \  ,}
then $C_{E=0}^*$ is large.  
Hence for coefficients that are of a size one would expect 
from the scale 
of strong interactions
the effective coupling constant of the $N^4$-operator 
could be significantly larger
than expected, due to a delicate cancellation.
Such a precise cancellation is certainly not unnatural 
in the sense that in order to find a
large scattering length compared to any physical length 
scale a cancellation of some sort is
almost certainly required.
Also, note that for a large value of $C^{\Delta\Delta}$ 
we return to the familiar
scenario discussed by Weinberg, where the theory with 
$\Delta$'s is identical to the theory
without $\Delta$'s.
In fact, for reasonable values such as 
$C^{\Delta\Delta} \sim 1/(100 \ {\rm MeV})^2$ which gives
$C^{\Delta\Delta} \ G_{E=0}^{\Delta\Delta} ({\bf 0},{\bf 0}) 
\sim 5.5$ the 
theory behaves largely like the theory of
nucleons alone.

It is illuminating to consider special cases for the 
couplings that reproduce the scattering
length in this channel.
For  $C^{NN} = C^{\Delta N} = C^{\Delta\Delta} = C_0$ 
we find from \scattrange\ 
that 
\eqn\nopiCfix{\eqalign{
C_0 & = {4 \pi\over M_N} 
{1\over \left( {1\over a_0} + \sqrt{2 M_N\delta}\right) }
\cr
& \sim  +{1\over (234 {\rm MeV})^2}
}\ \ \ .}
The inclusion of the $\Delta$'s has introduced an 
intrinsic length scale $\sqrt{2 M_N\delta}$,
which now sets the scale for the coefficients of 
the four-baryon operators.
More importantly, we see that for a large scattering length, 
such as is found in this
scattering channel, the coefficients must be finely tuned so that 
there is substantial cancellation between 
$1/C_0$ and $G_{E=0}^{\Delta\Delta}({\bf 0},{\bf 0})$ 
in \leadnopi\ 
as discussed above.
The effective range for this parameter set is 
$r_{\rm eff.} \sim -0.26 {\rm fm}$,
which is to be compared to the measured value of 
$r_{\rm eff.} \sim +2.73 {\rm fm}$.
For another choice of coefficients, 
$C^{NN} = {1\over 2} C^{\Delta N} = 
{1\over 4} C^{\Delta\Delta} = C_0$,
a fit to the scattering length yields 
$C_0 \sim +{1\over (470 {\rm MeV})^2}$ and in turn 
we find the effective range to be
$r_{\rm eff.} \sim -1.05 {\rm fm}$.
We see in these two examples the problem of having 
large coefficients of the four-nucleon
operators is traded for a fine tuning problem when 
the $\Delta$'s are included.
An analogous situation would arise if one was to 
regulate the nucleon  theory alone with a
momentum cut-off 
procedure~\foot{We thank Peter Lepage for discussions on this point}
where the scattering length results from a fine 
tuning between the cutoff and coefficients in
the expansion.

It is simple to see from \leadnopi , \cstar\ and 
\scattrange\ how the $\Delta$'s decouple from
the theory as they become much more massive than the nucleon.
As $\delta\rightarrow\infty$ the green function 
$G_E^{\Delta\Delta}({\bf 0},{\bf 0})$ becomes large and we can
expand $C_E^*$ in inverse powers of  
$G_E^{\Delta\Delta}({\bf 0},{\bf 0})$,
\eqn\cstarexp{\eqalign{
C_E^* & = C^{NN} - { (C^{\Delta N})^2\over C^{\Delta\Delta}} 
- {4\pi\over M_N \sqrt{2 M_N\delta - |{\bf p}|^2 } } 
\left( {C^{\Delta N}\over
C^{\Delta\Delta}}\right)^2
\ +\ ...
}\ \ \ .}
This tends to a momentum independent constant 
as $\delta\rightarrow\infty$, and we recover
Weinbergs summation of nucleon bubbles, with a coefficient
\eqn\cdefwein{\eqalign{
C & =  C^{NN} - { (C^{\Delta N})^2\over C^{\Delta\Delta}} 
}\ \ \ .}
Expanding $p \cot ( \delta )$,  the scattering length 
and effective range in \scattrange\
become
\eqn\heavypar{\eqalign{
a & \rightarrow {M_N\over 4\pi}
\left[  C^{NN} - { (C^{\Delta N})^2\over C^{\Delta\Delta}}
\right]
 - \left( {C^{\Delta N}\over C^{\Delta\Delta}}\right)^2 
 { 1\over \sqrt{2 M_N\delta }  }
 \ +\ .....
 \cr
 r_{\rm eff} & \rightarrow 
 -\left({C^{\Delta N}\over C^{\Delta\Delta}}\right)^2 {1\over
 a^2}{1\over  (2 M_N\delta)^{3\over 2}}
 \ +\ .....
 }\ \ \ .}

One should note that \cdefwein\ demonstrates that the 
coupling of the $N^4$ operator in the
theory of nucleons alone
is not directly related to the coupling of the $N^4$ 
operator in the theory including
$\Delta$'s.   
This means that a direct comparision between extractions 
of $C_E^*$ in \cstar\ from nuclei or
NN interactions with predictions or relations from 
QCD  (e.g. \kapsavNC\ ) is not straightforward.

\bigskip
\newsec{The Real World: Non-Zero Axial Couplings.}

In the previous section we discussed a theory with 
vanishing axial couplings.
Although not directly related to nature, it did 
demonstrate interesting behaviour, particularly
in how the $\Delta$'s decouple from the nucleon sector.
In this section we include leading order pion exchange 
into the system, and extend the work of
\ksw .
The procedure for the inclusion of pions is essentially 
that discussed in \ksw\  with the
additional complication of including the $\Delta\Delta$ 
channel and so
we will only present an outline of the procedure.

The calculation of the phase shift in the $^1 S_0$ 
channel is split up into two parts.
The first part involves computing the amplitude from
the yukawa ladder sum arising from single pion exchange 
$i {\cal A}_\pi$
where there are both $NN$ and 
$\deldel$ intermediate states, see 
\fig\yukladd{The ladder sum of yukawa exchanges with 
both $NN$ and 
$\deldel$ intermediate states.}.
The second part is computing the amplitude from the
sum of all possible insertions of the local 
contact terms dressed with  yukawa ladder
exchanges 
$i {\cal A}_{\rm bub.}$, see
\fig\yukbubb{Insertions of the contact terms dressed 
with yukawa exchanges denoted by the light shaded regions.}.
The total amplitude is the sum of these two parts
\eqn\totamp{\eqalign{
i {\cal A} & = i {\cal A}_\pi + i {\cal A}_{\rm bub.}
}\ \ \ \  .}
To compute the yukawa ladder sum for $NN\rightarrow NN$ 
scattering one solves the
schrodinger equation for the wavefunctions $\Psi^N_E$ 
and $\Psi^\Delta_E $.
Defining $U^N = r \Psi^N_E $ and  $U^\Delta = r \Psi^\Delta_E $, 
we solve
\eqn\laddeqn{\eqalign{
\left[ -{1\over M_N} {d^2\over dr^2}
\left(\matrix{1&0\cr 0&1}\right)
- {e^{- m_\pi r}\over r} 
\left(\matrix{\alpha^{NN}&\alpha^{\Delta N}\cr 
\alpha^{\Delta N}&\alpha^{\Delta \Delta}}\right)
- \left(\matrix{E&0\cr 0&E-2\delta}\right)\right]
\left(\matrix{U^N\cr U^\Delta}\right) =0
}\ \ \ ,}
subject to the condition that 
$U^\Delta\rightarrow e^{-r \sqrt{2 M_N\delta-|{\bf p}|^2}}$
asymptotically (as we are working below the energy 
required for $\Delta\Delta$ production).   
The NN wavefunction obtained as a solution to \laddeqn\ 
is compared with the asymptotic
behaviour 
\eqn\asymptot{\eqalign{
\Psi^N_E (r)  \too {r\rightarrow\infty} &
-{i\over 2}\left( e^{2 i \delta_\pi} 
{e^{ipr}\over pr} - {e^{-ipr}\over pr} \right)
}\ \ ,}
for an s-wave, to obtain the phase shift from pion 
exchange alone, $\delta_\pi$.
The yukawa couplings in \laddeqn\ are related to the 
couplings appearing in \axiallag\ 
by
\eqn\yukaxial{\eqalign{
\alpha^{NN} & = {g_A^2 m_\pi^2\over 8 \pi f_\pi^2}
\cr
\alpha^{\Delta N} & = 
{\sqrt{5} (g^{\Delta N})^2 m_\pi^2\over 9 \pi f_\pi^2}
\cr
\alpha^{\Delta\Delta} & = 
{11 (g^{\Delta\Delta})^2 m_\pi^2\over 72 \pi f_\pi^2}
}\ \ \ \  .}
The amplitude $i {\cal A}_\pi$ can be found directly from the phase
shift
\eqn\phaseamp{\eqalign{
i {\cal A}_\pi & = {4\pi\over 2 M_N p}
\left( e^{2 i \delta_\pi} -1\right)
}\ \ \ .}

The amplitude resulting from multiple insertions of the 
contact operators can be found 
straightforwardly by explicit computation of the  bubble 
chains.
The coefficients that appear in the bubble chain are 
modified from those that appear in
\locallag\ due to a contribution from the part of the pion 
potential of the form
$\delta^3 ({\bf r})$.
As in \ksw\ we use a tilde to denote the modified coefficients
\eqn\cmodify{\eqalign{
\tilde C^{NN} & = C^{NN} + {g_A^2\over 2 f_\pi^2}
\cr
\tilde C^{\Delta N} & = C^{\Delta N} 
+ {20\over 9\sqrt{5}}{(g^{\Delta N})^2\over f_\pi^2}
\cr
\tilde C^{\Delta\Delta} & = C^{\Delta\Delta} 
+ {11\over 18} {(g^{\Delta\Delta})^2\over f_\pi^2}
}\ \ .}
Defining 
\eqn\bubdef{\eqalign{
X & = 1-
\tilde C^{NN} G_E^{NN, \msb}({\bf 0},{\bf 0})
\cr
Y & = 1- \tilde C^{\Delta\Delta} 
G_E^{\Delta\Delta, \msb}({\bf 0},{\bf 0})
\cr
Z & = \tilde C^{\Delta N} 
G_E^{\Delta\Delta, \msb} ({\bf 0},{\bf 0})
+ \tilde C^{N N} 
G_E^{\Delta N, \msb}({\bf 0},{\bf 0})
\cr
W & = \tilde C^{\Delta N} 
G_E^{NN, \msb} ({\bf 0},{\bf 0})
+ \tilde C^{\Delta\Delta} 
G_E^{\Delta N, \msb}({\bf 0},{\bf 0})
}\ \ \ ,}
the bubble sum is 
\eqn\bubblesum{\eqalign{
i A_{\rm bub.}  = 
- i & \left[ 
{\left[ \Psi_E^N (0) \right]^2\over X}
\left[
\tilde C^{NN}
 + 
\tilde C^{\Delta N}
{ Z \over X Y - Z W}
\right]
\right.
\cr
&\left.
\ +\  2\  { \Psi_E^N (0) \Psi_E^\Delta (0) 
\tilde C^{\Delta N} \over X Y - Z W}
\right.
\cr
&\left.
\ +\ {\left[ \Psi_E^\Delta (0) \right]^2\over Y}
\left[
\tilde C^{\Delta\Delta}
 + 
\tilde C^{\Delta N}
{ W \over X Y - Z W}
\right]\ \ 
\right]
}\ \ \ .}
The functions $\Psi_E^{N} (0)$ and $ \Psi_E^{\Delta} (0)$ 
are determined by 
\laddeqn\ and $G_E^{ij , \msb}({\bf 0},{\bf 0})$ 
are green functions  regulated in $\msb$.
We compute them numerically by finding solutions to
\eqn\contacteqn{\eqalign{
\left[ -{1\over M_N} \nabla^2\left(\matrix{1&0\cr 0&1}\right)
- {e^{- m_\pi r}\over r} 
\left(\matrix{\alpha^{NN}&\alpha^{\Delta N}\cr 
\alpha^{\Delta N}&\alpha^{\Delta \Delta}}\right)
- \left(\matrix{E&0\cr 0&E-2\delta}\right)\right]
\left(\matrix{G_E^{NN}\cr G_E^{N\Delta}}\right) 
= & \delta^3 ({\bf r}) \left(\matrix{1&0\cr
0&0}\right)
\cr
\left[ -{1\over M_N} \nabla^2\left(\matrix{1&0\cr 0&1}\right)
- {e^{- m_\pi r}\over r} 
\left(\matrix{\alpha^{NN}&\alpha^{\Delta N}\cr 
\alpha^{\Delta N}&\alpha^{\Delta \Delta}}\right)
- \left(\matrix{E&0\cr 0&E-2\delta}\right)\right]
\left(\matrix{G_E^{\Delta N}\cr G_E^{\Delta\Delta}}\right) = & 
\delta^3 ({\bf r}) \left(\matrix{0&0\cr 0&1}\right)
}\ \ \ .}
The boundary conditions are that 
$G_E^{NN}$ and $G_E^{\Delta N}$  have no incoming 
wave and that 
$G_E^{N\Delta}$ and $G_E^{\Delta \Delta}$
fall exponentially at large radius.
We have used shorthand for 
$G_E^{ij} = G_E^{ij}({\bf r},{\bf 0})$.
The  $G_E^{ij}({\bf 0},{\bf 0})$  are divergent as computed from 
\contacteqn\  and are 
regulated using dimensional regularization with modified 
minimal subtraction $\msb$, in
analogy with the procedure of \ksw .
One finds that the green functions defined in $\msb$ are 
simply related to those determined in
\contacteqn\ by
\eqn\greemmsb{\eqalign{
G_E^{ij, \msb}({\bf 0},{\bf 0}) 
& = \lim_{r\to 0}
\left[   G_E^{ij}({\bf r},{\bf 0}) + {M_N\over 4\pi r}\delta^{ij} 
-{\alpha^{ij} M_N^2\over 8\pi}\left[2\log(\mu r) 
+ 2\gamma - 1 \right]
\right]
}\ \ \ .}
The constants $C^{ij}$ are defined in $\msb$ by this
procedure and must be fit to data.

We have the leading order amplitude defined in terms of the 
three  coefficients of the local
contact terms 
$C^{ij}$ and the three axial coupling constants $g^{ij}$.
As mentioned previously, the axial couplings constants are 
found to obey the SU(4) relations
rather well and we will use these relations for the 
remaining discussion.
We discuss results obtained for some
choices of the coefficients of the contact terms
 but are unable to say anything 
more concrete.
By contruction,
the low momentum behaviour of the phase shift is 
the same for all choices of $C^{ij}$ that
reproduce the scattering length, whose structure is 
dominated at very low momentum by 
$\cot (\delta) = -1/(k a_0)$.
One finds that deviations in the phase shift between 
different choices of $C^{ij}$ arise around $15 {\rm MeV}$.   
This comes as no surprise as we know that
a higher dimension operator with a scale
of $\sim 100{\rm MeV}$ is required to reproduce the 
effective range in the absence of $\Delta$'s
and that, in general,  the $\Delta$'s themselves do not 
reproduce the effective range.

Lets us again consider the case of 
$C^{NN} = C^{\Delta N} = C^{\Delta\Delta} = C_0$ 
where fitting to
the scattering length requires 
$C_0 = -1/(106.2 {\rm MeV})^2$.
The effective range predicted by this choice of parameters is 
$r_{\rm eff.} = 4.3 {\rm fm}$.
However, despite this reasonably good fit to these two low energy 
parameters, the phase shift over a larger
momentum interval, below $|{\bf p}| < 150 {\rm MeV}$ is 
relatively poor.
The same can be said for a choice of parameters that
reproduces both the scattering length and the effective range,
$C^{\Delta\Delta}/1.17 =C^{\Delta N}/1.17 
= C^{NN} = -1/( 110.995 {\rm MeV})^2$, 
as can be seen in 
\fig\CCCplot{The $^1S_0$ phase shift at leading order in the 
expansion verses centre-of-mass momentum.
The dashed curve is the the theory without $\Delta$'s and 
the solid curve is 
$C^{NN} = C^{\Delta N}/1.17 = 
C^{\Delta\Delta}/1.17 \sim -1/(110 {\rm MeV})^2$
fit to reproduce the scattering length and effective range.
The points correspond to the phase shift data from the 
Nijmegan partial wave analysis
\ref\nij{Nijmegan.}
}.
In fact, we find that the theory with $\Delta$'s is little 
different from the theory without 
$\Delta$'s, and that they have effectively decoupled.
The inclusion of the $\Delta$'s does not 
really improve the situation for
reasonable size four-baryon operators.

A discussion of an "optimal" expansion to 
perform in order to reproduce the
observed phase shift can be found in \ksw\ .
An expansion of $k \cot (\delta)$ in insertions of higher 
dimension operators 
was found to be more rapidly converging  than 
an expansion of the amplitude itself.
While such a procedure is simple for the inclusion of higher 
dimension $N^4$ operators, it is not
clear how to perform this for the inclusion of the $\Delta$'s, 
and other baryon resonances.
There is no expansion parameter to expand in when the 
$\Delta$'s and other resonances are
included at lowest order.  
Therefore, we should have expected all along to be able 
to do little
better than we were able to do in the theory with 
nucleons alone.
This is precisely what we found.
Recently, it was suggested 
\ref\kapdib{D. Kaplan, nucl-th/9610052 (1996).}
that the inclusion of an explicit dibaryon field in 
channels with large scattering lengths
allows one to recover `` reasonable'' power counting.
In such a scheme the effects of the $\deldel$ interactions could be 
included perturbatively and could be seen to decouple in a 
more tranparent manor.

\newsec{Conclusions}

We have included the $\Delta\Delta$ intermediate state into 
the calculation of NN scattering in
the $^1 S_0$ channel using the methods of \weinberg\ and \ksw\ .
In the limit of vanishing axial couplings of the pions to 
the $N$'s and $\Delta$'s we find that
the large scattering length can arise from coefficients of natural size 
(scale set by $\sqrt{2 M_N \delta}$) but a fine tuning is required.
Further, we can explicitly see how the $\Delta$'s decouple 
from the theory as either
the $\Delta-N$ mass difference becomes large or as 
the $\Delta^4$ interactions become
strong.  For realistic values of the contact terms (presently unknown) 
the leading order $NN$
scattering amplitude in the theory with $\Delta$'s closely resembles 
that of a theory with no $\Delta$'s.

When the axial couplings are reinstated one finds that the 
large scattering length is
reproduced by contact terms with a scale $\sim 100\  {\rm MeV}$.
One also can obtain effective ranges of the correct size for 
particular values of the $C^{ij}$ without the inclusion of 
higher dimension operators.
Generally, one fails to reproduce the phase shift over larger
momentum intervals at leading order in exactly the same way 
one fails to do so with nucleons
alone as the $\deldel$ state has effectively decoupled from the 
theory.
Its leading effect is to modify the value of the coefficients of the
four-nucleon contact terms.
A higher order contact term is still required to extend the range
of momentum over which the effective theory reproduces the data.

The coefficients of the local operators are not
presently determined by experiment, although two combinations 
are constrained  at leading order
by the scattering length and effective range 
(assuming SU(4) symmetry for the axial couplings).
This ignorance severely restricts the depth to which we can 
investigate the role of $\Delta$'s
in $NN$ scattering.

As we mentioned previously, the $\Delta\Delta$ intermediate 
state is not the lightest excited
state(s) that can appear in this channel.  
However,  the fact that the $\Delta$ and the $N$ together 
form an SU(4) multiplet 
leads one to suspect that the interactions between the 
$\Delta$'s and the $N$'s will be
stronger than between these baryons and baryons outside 
the SU(4) multiplet.
Such an effect is evidenced in the coupling between pions 
and nucleons with the $\Delta$ and
with other higher mass resonances such as the $N^*(1440)$.
It is for this reason that we have examined the 
$\Delta\Delta$ intermediate state.
However, we have learned from the limit of vanishing 
axial couplings that the stronger the
interaction, the less effect the state has on the 
overall scattering, i.e.
in the limit of large
$C^{\Delta\Delta}$ we return to the theory of $NN$ 
interactions alone at leading order.
It may turn out to be the case that the other 
intermediate states have a larger effect on the
scattering because they have a weaker interaction.

\bigskip

\newsec{Acknowledgements}

I would like to thank David Kaplan, Bira van Kolck, 
Gerry Miller, Peter Lepage  and Mark Wise
for useful discussions.  
I would also like to acknowledge the Institute for 
Nuclear Theory at the University of Washington
in Seattle for support and kind hospitality while 
this work was performed.
I acknowledge support by the Department of Energy under 
Grant No. DE-FG02-91-ER40682.
\listrefs
\listfigs

\centerline{Fig.~1}
\vskip 1.75in
\centerline{\epsfxsize=6in\epsfbox{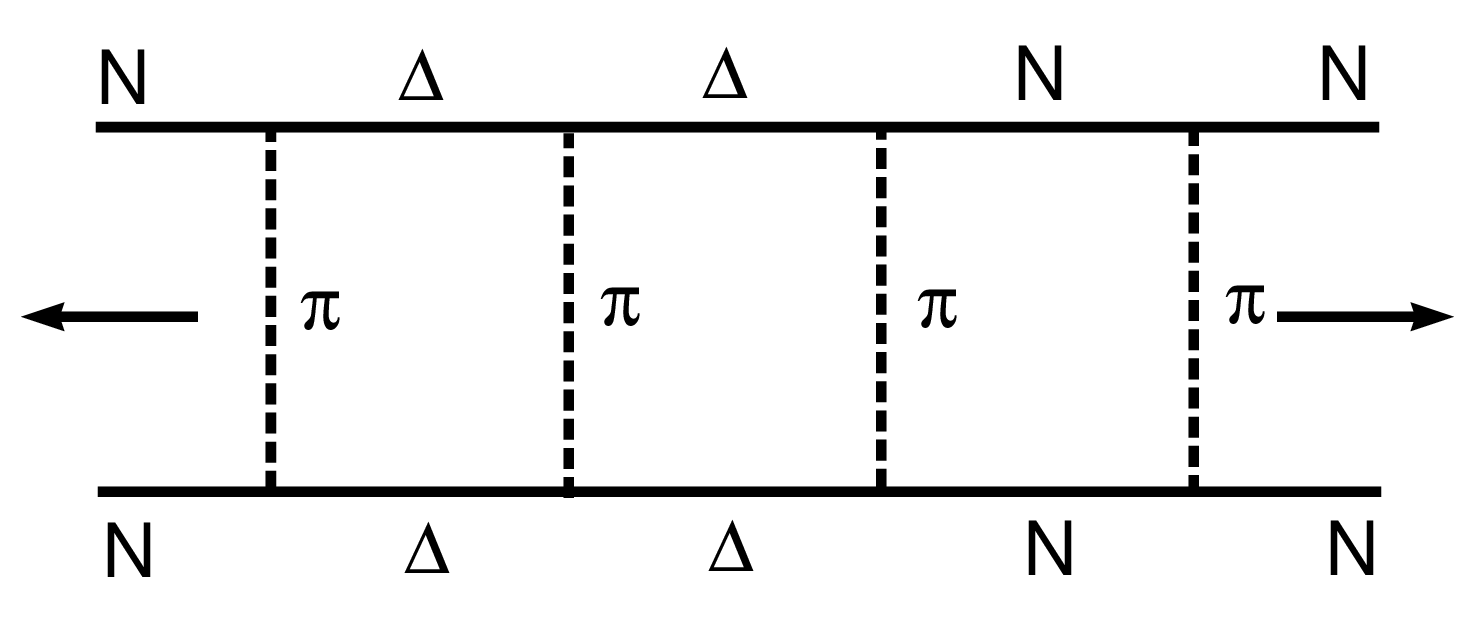}}
\vfill
\eject
\centerline{Fig.~2}
\vskip 1.75in
\centerline{\epsfxsize=6in\epsfbox{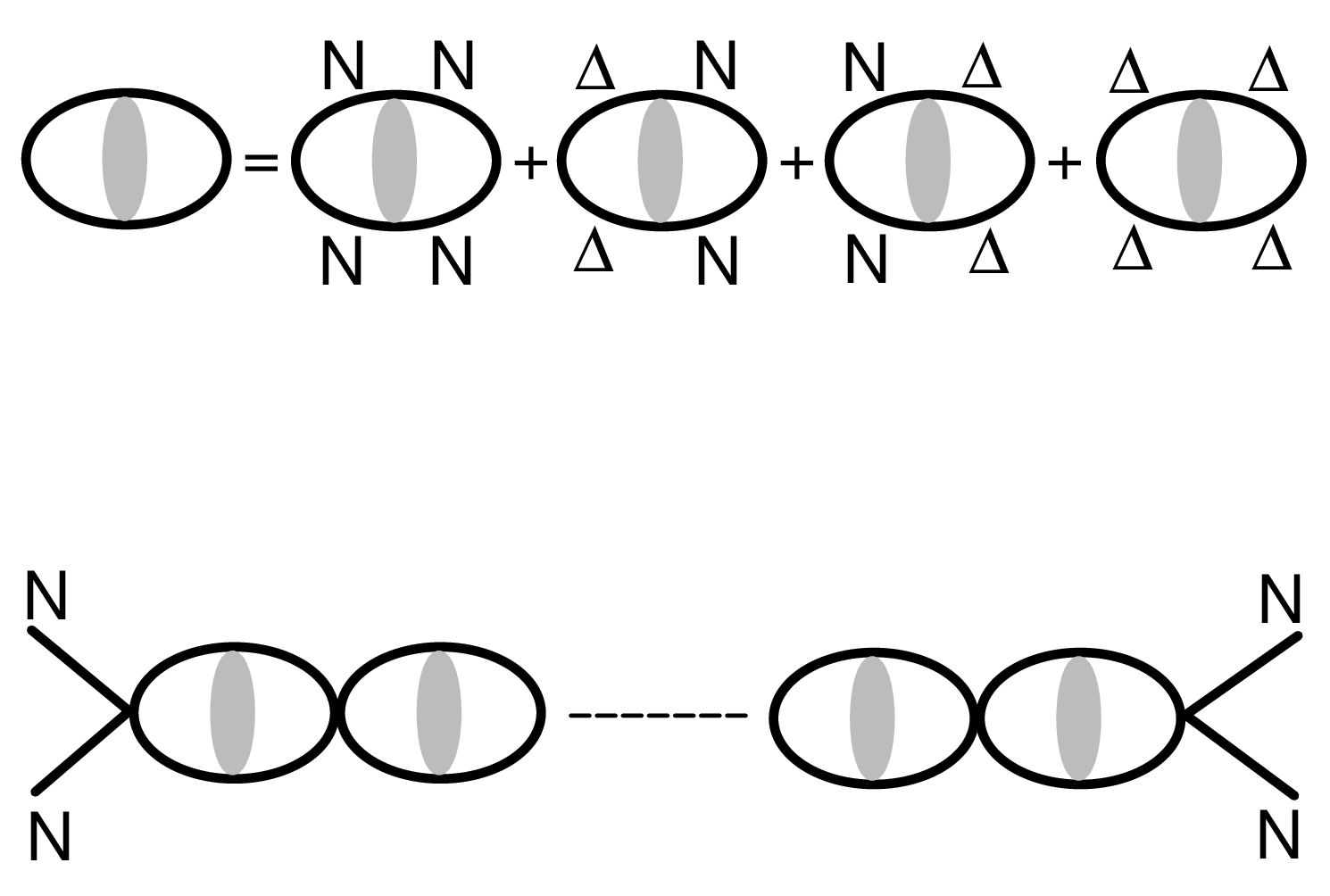}}
\vfill
\eject
\centerline{Fig.~3}
\vskip 1.75in
\centerline{\epsfxsize=6in\epsfbox{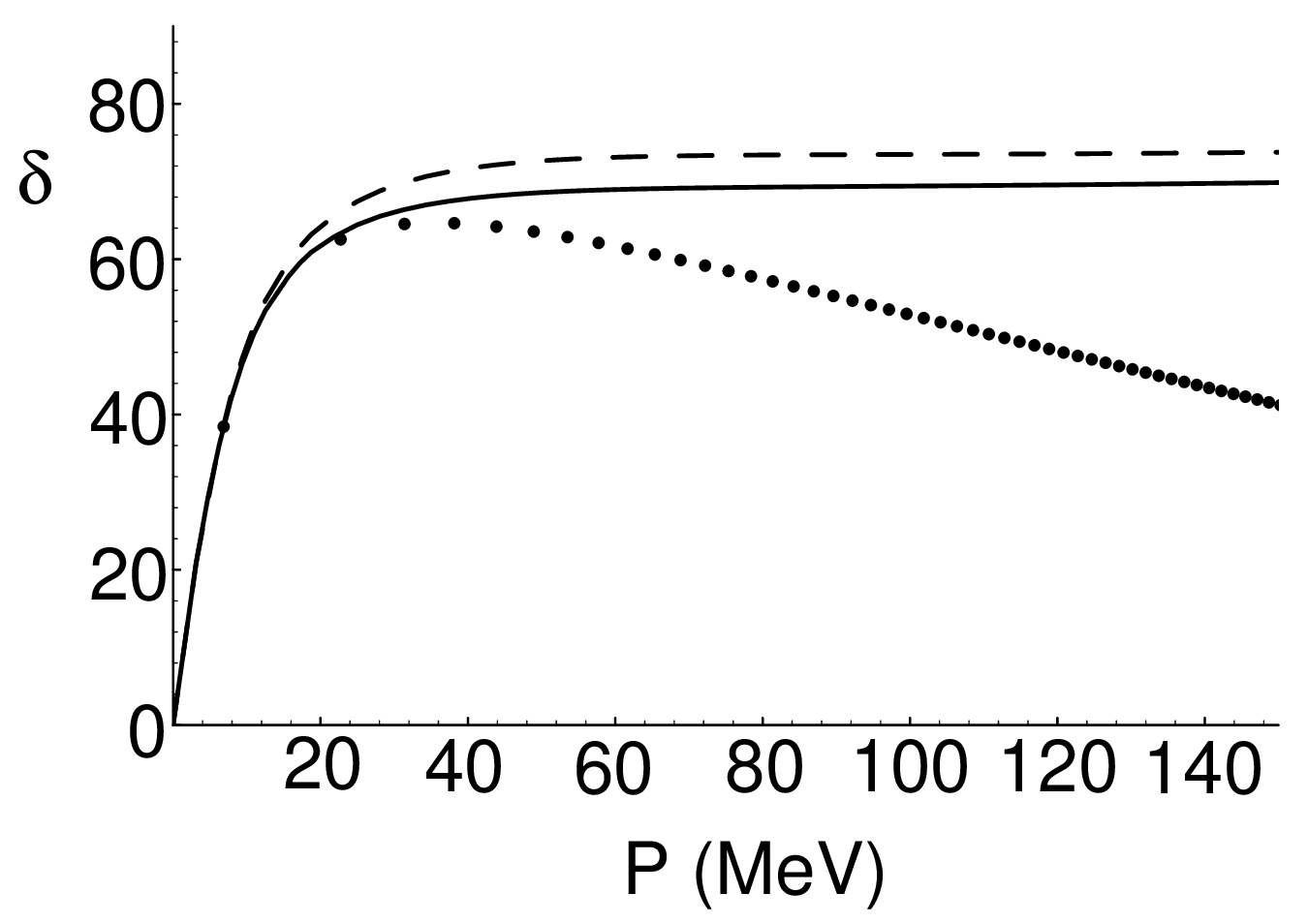}}
\vfill
\eject

\bye